# Exploring AI Futures Through Fictional News Articles

**Authors:** Martin Lindstam, Uppsala University; Elin Sporrong, Stockholm University; Camilo Sanchez, Aalto University; Petra Jääskeläinen, KTH Royal Institute of Technology

## Abstract

The aim of this workshop was to enable critical discussion on AI futures using fictional news articles and discussion groups. By collaboratively imagining and presenting future scenarios in a journalistic news article format, participants explored the socio-political, ethical and sustainability factors of AI through an accessible narrative form. Participants engaged in further anticipatory work by analyzing the issues raised by the articles in a group discussion, emphasizing the underlying motivations, assumptions and expectations conveyed within the news articles.

## Introduction

There has been an increasing attention to discussing ethics, sustainability, and politics of AI technologies. Along with the increasing urgency of discussing the consequences of AI technologies, there has also been a surge of interest in Design Fiction during the last 10 years (Bleecker, 2009; Tanenbaum et al., 2012; Dunne, 2013; Baumer, 2014; Penzenstadler, 2014; Blythe, 2014). This workshop made use of *fictional news articles* to materialize our imaginaries on AI and discuss how it will affect and change society now and in the future. By focusing on the *ethics, sustainability,* and *politics of AI futures*, we aimed to understand the potential future consequences of AI ten years from now – in 2034 — and how we can engage in the responsible design of AI.

This workshop was based on ICCC '23 workshop on Fictional Abstracts ("Workshops - ICCC 2023", 2023). As facilitators, we wanted to engage researchers and practitioners who are interested in the ethical reflection of emerging AI technology and practices. To foster a wide-ranging and critical discourse, we encouraged participants from different disciplines and backgrounds to bring their unique perspectives to the table. The workshop sought to offer a contribution by developing sensitivities on how to move forward in the responsible design and use of AI technologies.

The outcomes of the workshops showed diverse ways in which participants engaged with fictional news articles. While one group focused on extrapolating current trends – self-tracking and healthy lifestyles – into a dystopian speculative scenario of an AI-enabled lunar rave, the other presented emergent uses of AI by describing the development of an AI-driven food system in a rural community. Many discussions emerged during the workshop regarding themes, such as "world-building" and how power was distributed in the future scenarios. Another point for discussion was how perspectives are projected in the news articles. While the articles represented the authors' worldviews, readers themselves bring in their own perspectives to make sense of the speculation.

In the subsequent part of this paper, we will provide an account written by two of the workshop participants, reflecting on their experiences of the workshop, and the method of fictional news articles.

## Reflections from two workshop participants

Speculative methods invite people to engage with various temporal frames, what-ifs and counterfactual scenarios, to generate debate and new sociotechnical visions (Meskus & Tikka, 2022). While such creative engagements can spark inspiration and hope, we entered the workshop with a critical perspective drawing from our backgrounds in journalism, foresight and history of ideas.

Working with speculative methods, Elin has incorporated critical perspectives into her research on anticipations of AI in higher education, building on e.g. the discipline of anticipation, which provides tools for unpacking and expanding visions of the future (Miller et al, 2018). By identifying narratives and visions of AI technologies and the automation of teachers' practices, Elin's work aims to provide insights into how teachers' anticipations of AI, their assumptions and their values framing the future are interconnected in the context of higher education.

Coming from a journalistic background, Martin entered the workshop enmeshed in the emergent field of existential media studies (Lagerkvist, 2020), which seeks to recenter the often sidestepped and vulnerable human in visions of AI futures. The field of existential media studies emphasizes the future as a horizon of possibility. Building on this view of the future, Martin is working with Berardi's concept of futurability; denoting the idea that the horizon of possibility exists beyond entrenched contemporary norms and structures (Berardi, 2017). In his research, Martin is interested in techno-existential perspectives on sociotechnical imaginaries (Jasanoff & Kim, 2015), mainly concerning implementations of AI in the workplace.

With these backgrounds, both workshop participants attended the workshop with an understanding of news situated in a Swedish context, in which journalism is informed by a specific set of nationally agreed upon guidelines (Journalistförbundet, 2021) and media is commonly referred to as a third public power (Riksdagen, 2023). By these understandings of journalism, news writing is a narrative practice of compressing large amounts of information into intelligible text (Häggbom, 2020), structured around an understanding of public interests. As such, journalism is "society's conversation with itself", and newsworthiness is normatively informed (Häger, 2021). The prospect of news articles as venues for anticipation is interesting as the news article format is entangled in the web of journalistic moral and ideological practices (Gardeström, 2020), norms and mission statements.

## Producing the speculative artifact

Inspired by the format of fictional news articles, we considered how current narratives, values and assumptions regarding AI in society might be brought into our creative process of envisioning future trajectories, and how we could critically engage with our own anticipatory processes as our speculative visions unfolded in the workshop. In particular, we critically considered: i. newsworthy vs ordinary futures; ii. grand narratives of AI and; iii. fallacies in futuring for broadening our own thinking and imagination.

### i. Newsworthy vs ordinary futures

We considered that news reportages have the ability to transport media consumers to places they have never been, and to meet people they otherwise never would have met (Häggbom, 2020). Additionally, news reportages note the extraordinary – the newsworthy – from a normative standpoint of the journalist or newsroom, in response to people's interests in and of the surrounding world. As such, news as a speculative format may illuminate people's reflections on the newsworthy and ordinary. To imagine new places and the extraordinary, we first had to consider the familiar and the ordinary.

11

### ii. Grand narratives of AI

We contemplated domains of ordinary life and society within our lifeworlds. Being both teachers and students in higher education, we turned to the context of education that currently reflects the international political competition in technological progression, enacted through various AI strategies and policies (Bareis and Katzenbach, 2021). Situated in Western academia in 2023, our discursive understanding of ordinary education and future was discussed by critically noting current dominant narratives of AI in education, understood as an inevitable and necessary solution in education (Rahm and Kaun, 2022; Facer and Selwyn, 2021), echoed in our everyday contexts. Such narratives may marginalize alternative educational visions that draw on emerging perspectives in and on the world (Sporrong, 2024). Building on this understanding of the role of narratives, speculative methods may be more or less open to emerging perspectives by their capacity to both critically unpack and reproduce narratives that displace emerging thoughts and ideas (Ross, 2022).

### iii. Fallacies in futuring for broadening our own thinking and imagination

Speculation may draw on common errors in reasoning about the future that can hinder critical speculation (Dorr, 2017). Building on Dorr (2017), we considered ways to engage with the complexity of change and to imagine the future beyond the horizon set through an extension of the past. Thus, to critically consider dominant narratives and disentangle ourselves from fallacious prospective thinking, we tried to de-center Western tech utopias in our speculative artifacts and to re-imagine what education might mean in the future.

In this process, we situated ourselves in an imagined future in which education focused on sociality and wellbeing, rather than efficiency. Considering the environmental crises as an intertwined factor with future wellbeing, and thereby with educational development, we discussed how promises of "more efficient" education and environmental costs of technology stand in friction with environmental and human existential needs and the survival of living beings. As such, we problematized current projections of better education by AI and considered what AI could bring to education should the environment be a central and intertwined part of everyday school. Through this critical stance, we moved toward ideas of futures articulated by Berardi (2017), who underscores the importance of disentangling from entrenched norms and structures in order to perceive the horizon of possibility.

## Producing the speculative artifact

In our speculative project we set out to construct a rich and nuanced narrative as encouraged by the workshop facilitators. We began by situating education within a fictive rural municipality in which the community, learning, people's wellbeing and the environment were tightly connected aspects of ordinary life. In this vision, the educational and sustainability goals were interrelated for example, by schools working towards self-sufficiency through farming, situating schools within the broader context of environmental sustainability. By framing schools as central actors in common efforts to live more sustainably, we envisioned EarthTrack, a monitoring technology for streamlining agriculture that had been attuned to the need in private and public schools and their roles in the community. Two stakeholder perspectives were presented in the fictional news article to illustrate a tension and power dynamic in the vision: the perspective taken by a tech-positive local politician and that of a more cautious school principal. Supplying the final lines of the article, the principal is quoted: "I think we are seeing, particularly in the West, the many problems that schools have encountered because they were too quick to rush into this. That we have the possibility to do the same does not mean that we should".

By engaging in non-dominant visions of slow and alternative AI-implementation in education within the news article format, we were, to some extent, able to consider the complexity of the speculative scenario we imagined.

## Discussion and conclusion

The fictional news article method as outlined in the workshop is expressly reflective by viewing the future as an open space where multiple possible futures may be enacted. Berardi (2017) emphasizes the importance of this reflective approach not just in thinking about the future, but also humans' place in it. In line with Dorr (2017), Berardi (2017) describes and laments a reduction of the possible to the probable, and the probable to the necessary. Through our discussions, it became clear that we sought to generate expansive visions of the future by taking an awareness of the fallacious and reductive envisioning of the future.

Being prompted with multiple reflective questions, such as: "Are there any elements, contexts or situations from the article that we can already notice today?", and asked to reflect on the social, cultural, and environmental values present in our article, we found that many assumptions about education and currently guiding values were embedded in our vision. Despite our attempts to disentangle ourselves from dominant narratives, we identified extrapolative tendencies in our speculative artifact, reflected in the imagined future education system and ways of living in the rural municipality. For instance, physical schools were envisioned as a constant in the future, situated in or in conflict with a political discourse. Additionally, our community-oriented ideas that were central to our vision of sustainability drew on current ideas of communality - municipal private and public physical schools situated within a rural area where people lived together. Such ideas excluded alternative visions of schooling, nomad lifestyles and environmental tipping points that could completely alter our current lifestyles.

Considering such excluded visions, we note that AI visions may be speculative by various degrees and dependent on the materiality and normativity of the speculative method applied (Cerratto Pargman et al., 2023). For example, fictional news articles require multiple instances or sequences of speculation whereby we imagine not only the newsworthy event itself, but the actors, institutions and values that configure it as such – a complex web of moving parts, mediating our speculation. The materiality of the news article genre applied in the workshop was instantiated by specific stylistics and news-formats such as opinion pieces, interviews etc. By constructing visions by such material choices, we identified a further exclusion of alternative visions of journalism. Journalism became a lens for speculation rather than a practice, somewhat distanced from the logics and structures that commonly shape the news. As a lens, we found that journalism both challenged us and presented us with opportunities for speculation.

Understanding journalism as society's conversation with itself, we reflected on the nature of such a conversation in a fictive world. Journalism, ideally, focuses on human stories that motivate the newsworthiness of events by their importance in people's lives (Häger, 2021). Re-centering the human is a core concern of existential media studies, and we note that the fictional news article workshop method lends itself well to discussions of what it means to be human in relation to technology. As a closing session in the workshop, we discussed our intent with the construction of the speculative news article, noting that news carries with it normative assumptions of what is important information for the public. As such we found that in our efforts to adhere to the news article format, we were inadvertently making normative statements about what is important public information.

13

Finally, we consider that any process of disentanglement of dominant narratives may further substantiate, or reinforce them. For instance, we identify that efforts to disentangle from dominant narratives may be "reactionary", by arranging and structuring visions in response to visions under critique. In this workshop context of speculation, we identify that the collective and reflective closing discussion sparked new ideas and critical perspectives on anticipatory processes taking place, and what participants "bring into the future".

# References

Baumer, E.P., Ahn, J., Bie, M., Bonsignore, B., Borutecene, A., Buruk, O.T., Clegg, T., Druin, A., Echtler, F., Gruen, D., Guha, M. L., Hordatt, C., Kruger, A., Maidenbaum, S., Malu, M., McNally, B., Muller, M., Norooz, L., Norton, J., Özcan, O., Patterson, D., Riener, A., Ross, S., Rust, K., Schoening, J., Silberman, M. S., Tomlinson, B., and Yip, J. (2014). "CHI 2039: Speculative Research Visions". In Proc CHI '14. ACM, 761-770.

Berardi, F. (2017). Futurability: The Age of Impotence and the Horizon of Possibility. Verso books.

Bleecker, J. (2009). Design Fiction: A short essay on design, science, fact and fiction. Near Future Laboratory, 29.

Blythe, M. (2014). Research through design fiction: narrative in real and imaginary abstracts. In Proceedings of the SIGCHI Conference on Human Factors in Computing Systems (pp. 703-712). ACM.

Cerratto Pargman, T., Lindberg, Y. & Buch, A. Automation Is Coming! Exploring Future(s)-Oriented Methods in Education. Postdigit Sci Educ 5, 171–194 (2023). https://doi.org/10.1007/s42438-022-00349-6

Dorr, A. (2017). Common Errors in Reasoning about the Future: Three informal fallacies. Technological Forecasting & Social Change, 116. 322–330. https://doi.org/10.1016/j.techfore.2016.06.018.

Dunne, A., & Raby, F. (2013). Speculative everything: design, fiction, and social dreaming. MIT Press.

Facer, K., & Selwyn, N. (2021). Digital Technology and the Futures of Education: Towards 'Non-stupid' Optimism. UNESCO. https://unesdoc.unesco.org/ark:/48223/pf0000377071.

Gardeström, E. (2020). Vad är en journalist? In Gardeström, E., & Rehnberg, H. (Ed.) Vad är journalistik? 17-16.

Häger, B. (2021). Reporter: En grundbok i journalistik (3rd ed.). Studentlitteratur.

Häggbom, M. (2020). Journalistik är att formge berättelser. In Gardeström, E., & Rehnberg, H. (Ed.) Vad är journalistik? 61-65.

Workshops - ICCC 2023. (2023). Computational Creativity. https://computationalcreativity.net/iccc23/workshops/

Jasanoff, S., & Kim, S-H. (2015). Dreamscapes of Modernity: Sociotechnical Imaginaries and the Fabrication of Power. The University of Chicago Press

Journalistförbundet. (2021). Publicitetsregler. https://www.sjf.se/yrkesfragor/yrkesetik/spelregler-press-radio-och-tv/publicitetsregler

Lagerkvist, A. (2020). Existential Media: A Media Theory of the Limit Situation. Oxford University Press.

Meskus, M., & Tikka, E. (2024). Speculative approaches in social science and design research: Methodological implications of working in 'the gap' of uncertainty. Qualitative Research, 24(2), 209-228. https://doi.org/10.1177/14687941221129808

Miller, R., Poli, R. Rossel, P. (2018) The Discipline of Anticipation: Foundations for Futures Literacy In: Miller, R (Ed.) Transforming the Future: Anticipation in the 21st Century, (pp 48–60). Routledge

Penzenstadler, B., Tomlinson, B., Baumer, E., Pufal, M., Raturi, A., Richardson, D., Cakici, B., Chitchyan, R., Da Costa, B., Dombrowski, L., Picha Edwardsson, M., Eriksson, E., Franch, X., Hayes, G.R., Herzog, C., Lohmann, W., Mahaux, M., Mavin, A., Mazmanian, M., Nayebaziz, S., Norton, J., Pargman, D., Patterson, D.J., Pierson, J, Roher, K., Silberman, M.S., Simonson, K., Torrance, A.W., and van der Hoek, A. (2014). ICT4S 2029: What will be the systems supporting sustainability in 15 years. In Proceedings of the 2014 conference ICT for Sustainability (pp. 30-39). Atlantis Press.

Rahm, L., & Kaun, A. (2022). Imagining Mundane Automation: Historical Trajectories of Meaning Making Around Technological Change. In S. Pink, M. Berg, D. Lupton, & M. Ruckenstein (Eds.), Everyday Automation: Experiencing and Anticipating Emerging Technologies (pp. 23–43). London: Routledge. https://doi.org/10.4324/9781003170884-3.

Riksdagen. (2023). The Riksdag in Swedish society. https://www.riksdagen.se/en/how-the-riksdag-works/democracy/the-riksdag-in-swedish-society/

Ross, J. (2022). Digital Futures for Learning: Speculative Methods and Pedagogies (1st ed.). New York: Routledge. https://doi.org/10.4324/9781003202134.

Sporrong, E. (2024). Anticipation: Understanding the Role of Anticipation in Configuring Visions of Education. In Buch, A., Lindberg, Y., Cerratto Pargman, T. (eds). Framing Futures in Postdigital Education: Critical Concepts for Data-driven Practices. Springer

Tanenbaum, J., Tanenbaum, K., & Wakkary, R. (2012). Design fictions. In Proc. TEI'12. ACM, 347-350.